\documentclass[12pt]{article}

\usepackage[pdftex]{graphicx}

\usepackage{epsfig,cite}

\usepackage {amsmath}

\usepackage{color}
\usepackage{rotating}
\usepackage{amssymb}

\usepackage{slashed}

\include{epsf}

\newcount\timecount

\newcount\hours \newcount\minutes  \newcount\temp \newcount\pmhours

\hours = \time

\divide\hours by 60

\temp = \hours

\multiply\temp by 60

\minutes = \time

\advance\minutes by -\temp

\def\hour{\the\hours}

\def\minute{\ifnum\minutes<10 0\the\minutes

            \else\the\minutes\fi}

\def\clock{

\ifnum\hours=0 12:\minute\ AM

\else\ifnum\hours<12 \hour:\minute\ AM

      \else\ifnum\hours=12 12:\minute\ PM

            \else\ifnum\hours>12

                 \pmhours=\hours

                 \advance\pmhours by -12

                 \the\pmhours:\minute\ PM

                 \fi

            \fi

      \fi

\fi

}

\def\monthname{\relax\ifcase\month 0/\or January\or February\or

   March\or April\or May\or June\or July\or August\or September\or

   October\or November\or December\else\number\month/\fi}

\def\bold#1{\setbox0=\hbox{$#1$}%

     \kern-.025em\copy0\kern-\wd0

     \kern.05em\copy0\kern-\wd0

     \kern-.025em\raise.0433em\box0 }

\def\beq{\begin{equation}}

\def\eeq{\end{equation}}

\def\ga{\mathrel{\raise.3ex\hbox{$>$\kern-.75em\lower1ex\hbox{$\sim$}}}}

\def\la{\mathrel{\raise.3ex\hbox{$<$\kern-.75em\lower1ex\hbox{$\sim$}}}}

\def\gev{{\rm \, Ge\kern-0.125em V}}

\def\tev{{\rm \, Te\kern-0.125em V}}

\def\gyr{{\rm \, G\kern-0.125em yr}}

\def\gappeq{\mathrel{\rlap {\raise.5ex\hbox{$>$}}

{\lower.5ex\hbox{$\sim$}}}}

\def\lappeq{\mathrel{\rlap{\raise.5ex\hbox{$<$}}

{\lower.5ex\hbox{$\sim$}}}}

\def\Toprel#1\over#2{\mathrel{\mathop{#2}\limits^{#1}}}

\def\m12{m_{1\!/2}}

\def\bea{\begin{eqnarray}}

\def\eea{\end{eqnarray}}

\def\beqar{\begin{eqnarray}}

\def\eeqar{\end{eqnarray}}

\def\m{{\cal m}}

\begin{document}

\begin{titlepage}

\pagestyle{empty}

\baselineskip=21pt


\rightline{KCL-PH-TH/2013-10, LCTS/2013-05, CERN-PH-TH/2013-050}

\vskip 0.7in

\begin{center}

{\large {\bf Updated Global Analysis of Higgs Couplings}}

\end{center}

\begin{center}

\vskip 0.4in

 {\bf John~Ellis}$^{1,2}$
and {\bf Tevong~You}$^{1}$

\vskip 0.2in

{\small {\it

$^1${Theoretical Particle Physics and Cosmology Group, Physics Department, \\
King's College London, London WC2R 2LS, UK}\\

$^2${TH Division, Physics Department, CERN, CH-1211 Geneva 23, Switzerland}\\

}}

\vskip 0.4in

{\bf Abstract}

\end{center}

\baselineskip=18pt \noindent


{
There are many indirect and direct experimental indications that the new particle $H$
discovered by the ATLAS and CMS Collaborations has spin zero and (mostly)
positive parity, and that its couplings to other particles are correlated with their masses.
Beyond any reasonable doubt, it is a Higgs boson, and here we examine the extent to which its
couplings resemble those of the single Higgs boson of the Standard Model.
Our global analysis of its couplings to fermions and massive bosons determines that they
have the same relative sign as in the Standard Model. We also show directly that these couplings
are highly consistent with a dependence on particle masses that is linear to within a few \%, and scaled by
the conventional electroweak symmetry-breaking scale to within 10\%. We also give constraints
on loop-induced couplings, on the total Higgs decay width, and on possible invisible decays of the Higgs 
boson under various assumptions.
}


\vfill

\leftline{
March 2013}

\end{titlepage}

\baselineskip=18pt


\section{Introduction and Summary}

It has now been established with a high degree of confidence that
the new particle $H$ with mass $\sim 126$~GeV discovered by
the ATLAS~\cite{ATLASICHEP2012} and CMS~\cite{CMSICHEP2012first}
has spin zero and (mainly) positive-parity couplings, as expected for a Higgs boson~\cite{Higgs23}. Minimal spin-two
alternatives with graviton-like couplings have been disfavoured by
measurements of the $H$ couplings to vector bosons~\cite{ESY1}, and quite strongly excluded by
constraints on the energy dependence of $H$ production~\cite{ESY2}. The graviton-like
spin-two hypothesis has also been disfavoured strongly by analyses
of $H$ decays into $\gamma \gamma$~\cite{ATLASgaga}, $ZZ^*$ and $WW^*$ final states~\cite{Moriond2013,Moriond2013ATLAS},
and the positive-parity assignment is favoured by decays into $ZZ^*$,
in particular~\footnote{It is also impressive that the mass of the $H$ particle
coincides with the best fit for the mass of the Higgs boson found in a global 
fit to precision electroweak data taking account of pre-LHC searches at LEP 
and the TeVatron~\cite{GFitter}, and is also highly consistent with low-energy supersymmetry~\cite{susyH}.}.
Beyond any reasonable doubt, the $H$ particle is a
Higgs boson. 

In this paper we make updated global fits to the $H$
couplings to other particles with the aim of characterizing the extent
to which they resemble those of the Higgs boson of the Standard Model.
There has been considerable progress since our previous analysis of 
$H$ couplings~\cite{EY2}, including updates at the Hadron Collider Physics
conference in November 2012~\cite{HCPconf}, the CERN Council in December 2013~\cite{Jamboree},
the Moriond Electroweak Conference~\cite{MoriondEW} and the Aspen `Quo Vadis Higgs' 
Meeting in March 2013~\cite{Aspen}, and most recently 
an update of the CMS $H \to \gamma \gamma$ data at the Moriond QCD
session~\cite{MoriondQCD}.

There have been many analyses of the $H$ couplings~\cite{postdiscovery,EY2}, some also including the
Moriond 2013 data~\cite{postMoriond}. Many of these analyses, including those made by
the different experimental Collaborations, assume simple parameterizations
in which the couplings of the Standard Model Higgs boson to bosons and 
fermions are rescaled by factors $a_V$ and $c_f$, respectively (or
equivalently by factors $\kappa_{V,f}$)~\cite{Higgscouplings}. Fits with non-minimal couplings to
massive vector bosons have also been considered, as have fits in which
the loop-induced couplings to gluons and photons deviate by factors $c_{g,\gamma}$ 
from the values predicted in the Standard Model. The latter have been of
interest in view of the possible excess of $H \to \gamma \gamma$ decays
relative to the Standard Model prediction, particularly as reported by the
ATLAS Collaboration~\cite{ATLASgaga}. Since the $H\gamma\gamma$ coupling could in
principle receive contributions from new massive charged particles, and
the $Hgg$ coupling from new massive coloured particles, these are
particularly sensitive to new physics beyond the Standard Model.
In this paper we make updated global fits to the $H$ couplings within
such common phenomenological frameworks.

We also revisit parameterizations of the $H$ couplings to fermions and
bosons that were first considered in~\cite{EY2}, which are designed specifically
to probe the dependence of the $H$ couplings on particle masses. Namely,
we consider parameterizations of the $H$ couplings to fermions $\lambda_f$
and massive bosons $g_V$ of the form
 \beq
\lambda_f \; = \; \sqrt{2} \left(\frac{m_f}{M}\right)^{1 + \epsilon}, \; g_V \; = \; 2 \left(\frac{m_V^{2(1 + \epsilon)}}{M^{1 + 2\epsilon}}\right) ,
\label{scalingcouplings}
\eeq
which reduce to the couplings of the Standard Model Higgs boson in the double limit $\epsilon \to 0, M \to v = 246$~GeV.
This parameterization addresses explicitly the question the extent to which the $H$
particle resembles a quantum excitation~\cite{Higgs23} of the Englert-Brout-Higgs field that is
thought to give masses to the particles of the Standard Model~\cite{EB,Higgs1,Higgs23,GHK}.

We find that, in the absence of contributions from any particles beyond the Standard
Model, a combination of the Higgs signal strengths measured in different channels
is now very close to the Standard Model value, within 13\% at the 68\% CL. We also find,
for the first time, a strong preference for the couplings to bosons and fermions to have
the same sign, also as expected in the Standard Model, driven largely by the new CMS
result on $H \to \gamma \gamma$ decay. This also means that there is no significant
evidence of additional loop contributions to the $H \gamma \gamma$ beyond those
due to the top quark and the $W$ boson. Using the parameterization (\ref{scalingcouplings}),
we find that the dependence of the Higgs couplings to different particle species is within a few \%
of a linear dependence of their masses. Within the parameterization (\ref{scalingcouplings}),
or marginalizing over the $H$ couplings to Standard Model bosons and fermions, we find that
the total Higgs decay rate lies within 20\% of the Standard Model value at the 68\% CL.
If the couplings of the Higgs Boson to Standard Model
particles have their Standard Model values and there are no non-standard contributions to the
$H g g$ and $H \gamma \gamma$ amplitudes, the upper limit on invisible Higgs decays is 10\%
of the total Higgs decay rate.

\section{Summary of the Data}

The analysis of this paper is based mainly on the material presented by the LHC and TeVatron
experimental Collaborations at the March 2013 Moriond Conferences in La Thuile~\cite{MoriondEW,MoriondQCD}.
The following are some of the main features of interest among the new results:

\begin{itemize}

{\item The $H \to {\bar b}b$ signal strength reported by the TeVatron experiments has
reduced from $2.0 \pm 0.7$ to $1.6 \pm 0.75$ times the Standard Model value.}

{\item A new $H \to \tau^+ \tau^-$ result of $1.1 \pm 0.4$ has been reported by CMS, improving on the previous value of $0.7 \pm 0.5$.}

{\item The $H \to \gamma \gamma$ signal strength reported by ATLAS has reduced somewhat from
$1.80^{+0.4}_{-0.36}$ to $1.65^{+0.34}_{-0.30}$ times the Standard Model value. Most importantly, CMS has
reported a new result of $0.78^{+0.28}_{-0.26}$ for the signal strength using an MVA approach.}

{\item The $H \to WW^*$ signal strength reported by ATLAS has reduced from
$1.5 \pm 0.6$ to $1.01 \pm 0.31$ times the Standard Model value.}

\end{itemize}

All the latest available results from ATLAS, CMS and TeVatron are incorporated into our global fit. 
The experimental data are used to reconstruct the likelihood in a combination of three possible ways
according to the available information: 
1) using the official best-fit central value of $\mu$ with its 1-$\sigma$ error bars, 
2) using the given number of signal, background and observed events with their respective errors, 
or 3) reconstructing the central value of $\mu$ from the 95\% CL expected and observed $\mu$. Specifically,
the data inputs are as follows: 

\begin{itemize}

{\item The TeVatron $H \to {\bar b}b, \tau^+\tau^-, WW^*, \gamma\gamma$ combined best-fit $\mu$ and
1-$\sigma$ error bars from~\cite{TevatronMoriond2013}.} 

{\item The likelihood for the CMS 8 TeV $WW^*$ 0,1-jet analysis is
reconstructed from the numbers of events given in Table 4 of~\cite{CMSWWMoriond2013PAS}. 
The $WW^*$ 2-jet event numbers are instead taken from Table 3 of~\cite{HIG-12-042}. 
In addition, we use the fit values from~\cite{CMSICHEP2012} for the 7-TeV CMS $WW^*$ data. 
The ATLAS Collaboration provides 0,1-jet and 2-jet $\mu$ central values and 1-$\sigma$ ranges
for a combination of 7- and 8-TeV, which we treat effectively as 8 TeV. The percentages of the 
vector-boson fusion (VBF) production mode contributions
to the signals in the 0,1 and 2-jet channels are taken to be 2\%, 12\% and 81\%, respectively~\cite{ATLASlatestWWnote}. }

{\item For $H \to b{\bar b}$ in CMS we used the 7- and 8-TeV best-fit values from~\cite{CMSICHEP2012} and~\cite{CMSHCP2012}, 
while for ATLAS the likelihood was reconstructed from the 95\% CL expected and observed values of
$\mu$ at 7 and 8 TeV given in~\cite{HCP2012}. } 

{\item The CMS $H \to \tau^+\tau^-$ and $ZZ^*$ and $ZZ^*$ dijet rates were taken from the central values given in~\cite{Moriond2013}. 
Since no separate 7- and 8-TeV numbers are given for these, we treat them effectively as 8 TeV. 
Numbers of events for the ATLAS $H \to ZZ^*$ 7- and 8-TeV analyses are provided separately in~\cite{Moriond2013}, 
while the ATLAS $H \to \tau^+\tau^-$ likelihood is reconstructed using the 95\% expected and observed values of
$\mu$ given in~\cite{ATLAStautauHCP}. The VBF $\tau^+\tau^-$ efficiencies are taken from ~\cite{HIG-12-043}. }

{\item The CMS $\gamma\gamma$ central values are given for six (five) different 
subchannels at 8 (7) TeV in ~\cite{Moriond2013}, along with the 
percentage contributions from all production mechanisms in Table 2 in~\cite{CMSdiphotonICHEP2012}. The same information can be 
found for ATLAS at 7 TeV in~\cite{ATLASICHEP2012} and at 8 TeV in~\cite{Moriond2013}, 
broken down into eleven subchannels including two VBF-dominated ones. 
The CMS update is reported for a cut-based and MVA analysis; we use the MVA result, which has the greater sensitivity. }

\end{itemize}

For each individual experiment we have checked that our combinations of the likelihoods
for the various subchannels agree with official combinations with only slight exceptions, for example the CMS 7-TeV $\gamma\gamma$ analysis ($\mu=1.58^{+0.60}_{-0.61}$ instead of the official value 
of $1.69^{0.65}_{-0.59}$). When combined with the CMS 8-TeV data (for
which we reproduce the official central value) we calculate for the combined
CMS $\gamma\gamma$ data a value of $\mu=0.72^{+0.24}_{-0.26}$ 
(to be compared with the official value of $0.78^{+0.28}_{-0.26}$).
This difference of a fraction of the quoted error
does not impact significantly our overall results. 

\begin{figure}[h!]
\centering
\includegraphics[scale=0.5]{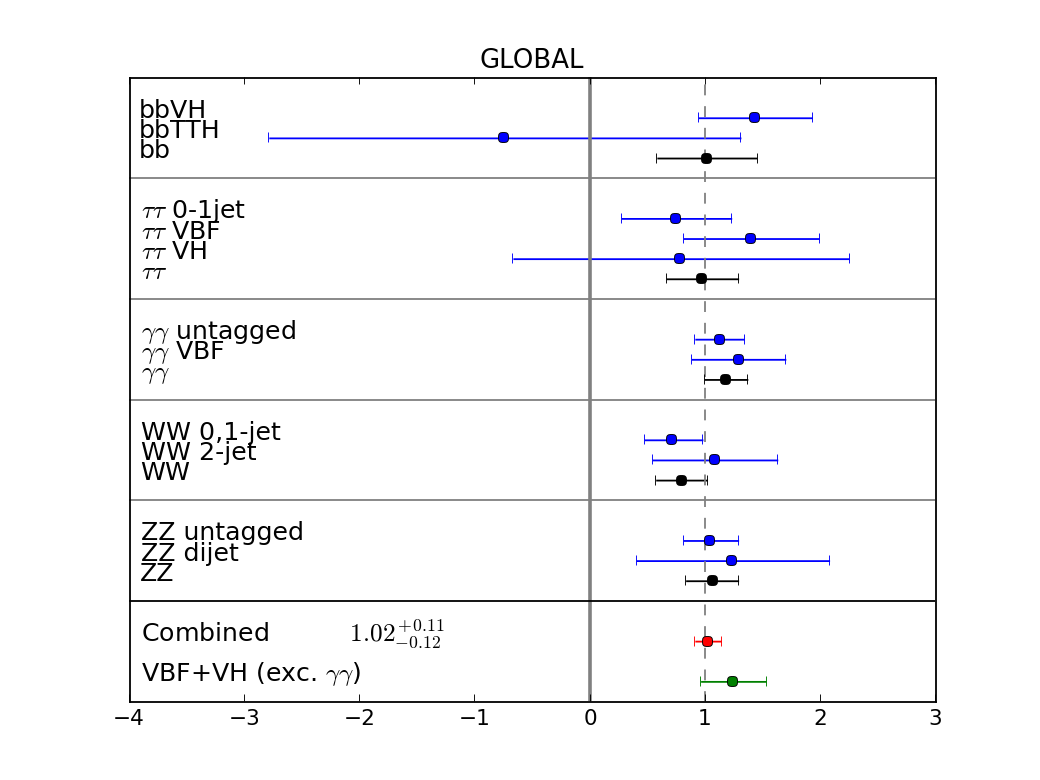}
\caption{\it A compilation of the Higgs signal strengths measured by the ATLAS, CDF, D0 and CMS
Collaborations in the ${\bar b} b$, $\tau^+ \tau^-$, $\gamma \gamma$, $WW^*$ and $ZZ^*$ final states.
We display the combinations of the different channels for each final state, and also the 
combination of all these measurements, with the result for the VBF and VH channels (excluding the $\gamma\gamma$ final state) shown separately in the bottom line.}
\label{fig:mulist}
\end{figure}

As a preliminary to our analysis, we compile in Fig.~\ref{fig:mulist} the overall signal strengths in the principal channels,
as calculated by combining the data from the different experiments. Thus, for example,
in the first line we report the $V + (H \to {\bar b} b)$ signal strength found by combining
the data on associated $V + H$ production from the TeVatron and LHC. As can be seen in the
second line, so far there is no significant indication of associated ${\bar t} t + H$ production.
The third line in Fig.~\ref{fig:mulist} combines the experimental information on the 
$H \to {\bar b} b$ signal strengths in these two channels.
Signals for $H \to \tau^+ \tau^-$ decay have now been reported in various production
channels, as reported in the next three lines of Fig.~\ref{fig:mulist}, and the
combined signal strength is given in the following line. As we have discussed, data are
available on $H \to \gamma \gamma$ final states following production in gluon-gluon
collisions and via vector-boson fusion. The central values of the corresponding
signal strengths are now only slightly larger than the Standard Model predictions, and we return
later to a discussion of the significance of these measurements. 
The signal strengths in the $H \to WW^*$ and $ZZ^*$ final states are very much
in line with the predictions of the Standard Model. These dominate the determination
of the combined signal strength reported in the last line of Fig.~\ref{fig:mulist}, together
with the $\gamma \gamma$ final state. It is striking that the available data already
constrain the combined Higgs signal strength to be very close to the Standard Model value:
\begin{equation}
\mu \; = \; 1.02^{+ 0.11}_{- 0.12} \, .
\label{mu}
\end{equation}
We present separately the
combined signal strength in the VBF and VH channels without the loop-induced $\gamma\gamma$ final state, 
which lies slightly (but not significantly)
above the Standard Model value. To the extent that a signal with direct Higgs couplings in both the initial and 
final state is established, this combination disfavours models that predict a universal suppression of the Higgs couplings~\footnote{We
address later  in a full fit of the effective couplings of the Higgs to photons and gluons 
the question whether an enhancement of the loop-induced gluon fusion production could compemsate for
this by contaminating the VBF cut selection.}.

\section{Higgs Couplings to Bosons and Fermions}

Our first step in analyzing the implications of these data uses
the following effective low-energy nonlinear Lagrangian for the
electroweak symmetry-breaking sector~\cite{ac}:
\bea
{\cal L}_{eff} \; & = & \; \frac{v^2}{4} {\rm Tr} \left(D_\mu U D^\mu U^\dagger \right) \times \left[ 1 + 2 a \frac{H}{v} +  \dots \right] \nonumber \\
& - & \frac{v}{\sqrt{2}} \Sigma_f {\bar f}_L\lambda_f f_R \left[ 1 + c_f \frac{H}{v} + \dots \right] + h.c. \, ,
\label{effL}
\eea
where $U$ is a unitary $2 \times 2$ matrix parametrizing the three Nambu-Goldstone fields
that give masses to the $W^\pm$ and $Z^0$ bosons, $H$ is the physical Higgs boson field and $v \sim 246$~GeV is the
conventional electroweak symmetry-breaking scale. The coefficients $\lambda_f$ are the Standard Model Yukawa couplings
of the fermion flavours $f$, and the factors $a$ and $c_f$ characterize the 
deviations from the Standard Model Higgs boson couplings of the $H$ couplings to massive
vector bosons and the fermions $f$, respectively. The couplings of the Higgs boson to massless boson pairs
$gg$ and $\gamma \gamma$ are described by the following dimension-5 loop-induced couplings:
\begin{equation}
{\cal L}_{\Delta} \; = \; - \left[ \frac{\alpha_s}{8 \pi} c_g b_g 
G_{a \mu \nu} G_a^{\mu \nu} + \frac{\alpha_{em}}{8 \pi} c_\gamma b_\gamma F_{\mu \nu} F^{\mu \nu} \right] \left(\frac{H}{V}\right) \, ,
\label{triangles}
\end{equation}
where the coefficients $b_{g, \gamma}$ are those found in the Standard Model, and the factors $c_{g, \gamma}$
characterize the deviations from the Standard Model predictions for the $H$ couplings to massless
vector bosons.

One specific model for a common rescaling factor of all fermion and vector boson Higgs couplings
is a minimal composite Higgs scenario~\cite{ac}, the MCHM4, in which the compositeness scale $f$ is related to $(a,c)$ by
\begin{equation*}
a = c = \sqrt{1-\left(\frac{v}{f}\right)^2} \, .
\end{equation*}
A similar universal suppression is found in pseudo-dilaton models.
A variant of this minimal model with a different embedding of the Standard Model fermions in SO(5) 
representations of the new strong sector, the MCHM5, has separate vector and fermion rescalings:
\begin{equation*}
a = \sqrt{1-\left(\frac{v}{f}\right)^2} 	\quad , \quad	 c = \frac{1-2\left(\frac{v}{f}\right)^2}{\sqrt{1-\left(\frac{v}{f}\right)^2}} 	\, .
\end{equation*}
In the following we confront the data with these specific models, as well as an `anti-dilaton' scenario in
which $c = - a$.

\begin{figure}[h!]
\centering
\includegraphics[scale=0.4]{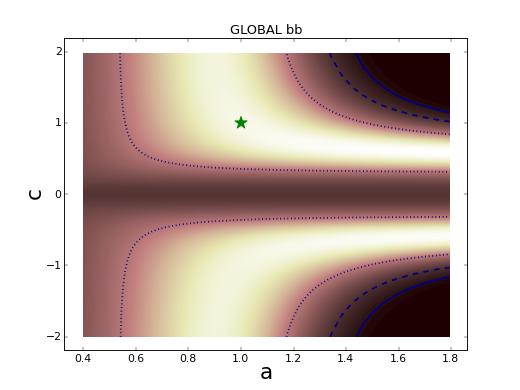}
\includegraphics[scale=0.4]{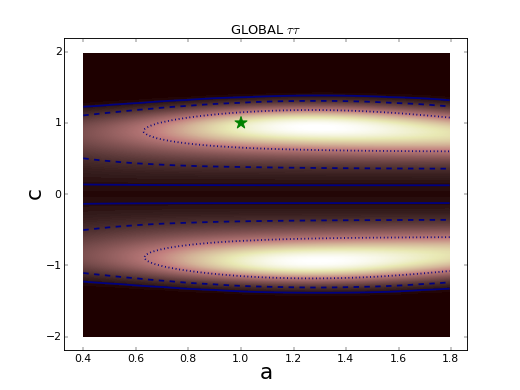}\\
\includegraphics[scale=0.4]{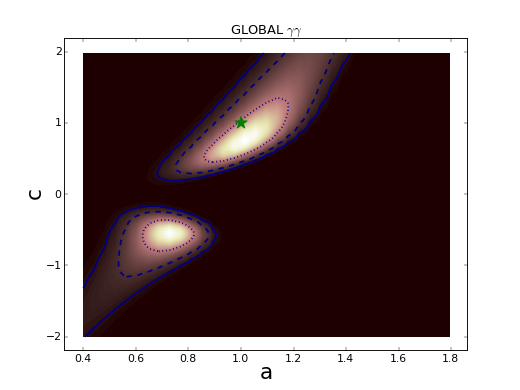}
\includegraphics[scale=0.4]{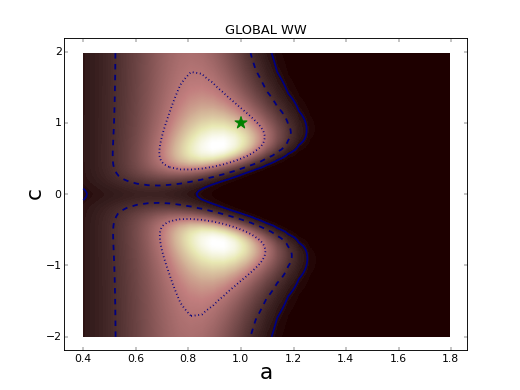}\\
\includegraphics[scale=0.4]{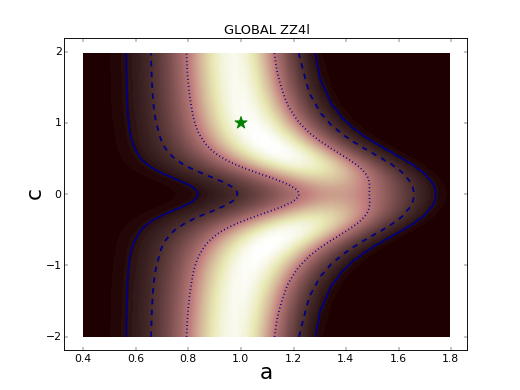}
\includegraphics[scale=0.4]{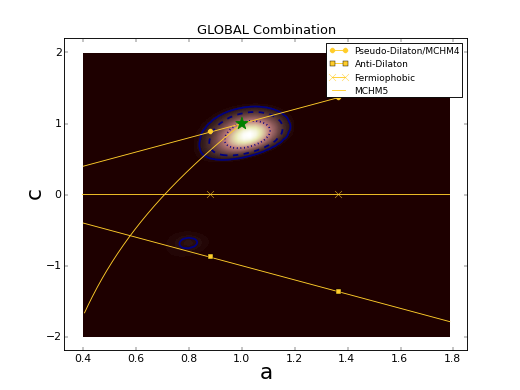}
\caption{\it The constraints in the $(a, c)$ plane imposed by the measurements in Fig.~\ref{fig:mulist}
 in the ${\bar b} b$ final state (top left), in the $\tau^+ \tau^-$ final state (top right),
 in the $\gamma \gamma$ final state (middle left), in the $WW^*$ final state (middle right)
 and in the $ZZ^*$ final state (bottom left). The combination of all these constraints is
 shown in the bottom right panel.}
\label{fig:ac}
\end{figure}

Fig.~\ref{fig:ac} compiles the constraints imposed by the data summarized in Fig.~\ref{fig:mulist}
on the factors $(a, c)$ in the effective Lagrangian (\ref{effL}), assuming universality in the
fermion factors $c_f \equiv c$, and assuming that no non-Standard-Model particles contribute to
the anomaly factors $c_{g, \gamma}$, which therefore are determined by a combination of the factors $c_t = c$
and $a_W = a$. In each panel of Fig.~\ref{fig:ac} and similar subsequent figures, the more likely regions of
parameter space have lighter shading, and the 68, 95 and 99\% CL contours are indicated by dotted,
dashed and solid lines, respectively.

We see again in the top row of panels of Fig.~\ref{fig:ac} that the data on $H \to {\bar b} b$ decays
(left) and $\tau^+ \tau^-$ decays (right) are entirely consistent with the Standard Model predictions
$(a, c) = (1, 1)$. The region of the $(a, c)$ plane favoured by the ${\bar b} b$ data manifests a
correlation between $a$ and $c$ that arises because the dominant production mechanism is
associated $V +X$ production, which is $\propto a^2$. On the other hand, the region of the $(a, c)$ 
plane favoured by the $\tau^+ \tau^-$ data exhibits a weaker correlation between $a$ and $c$,
reflecting the importance of data on production via gluon fusion in this case.
As was to be expected from the compilation in Fig.~\ref{fig:mulist},
the $\gamma \gamma$ data displayed in the middle left panel of Fig.~\ref{fig:ac}
are now compatible with the Standard Model prediction $(a, c) = (1, 1)$, following inclusion of the latest CMS result.
The data on $H \to WW^*$ (middle right panel of Fig.~\ref{fig:ac}) and $ZZ^*$ decays 
(bottom left panel) are also entirely consistent with $(a, c) = (1, 1)$.

We draw attention to the importance of the 2-jet analyses, which select a VBF-enriched sample,
in disfavouring bands of the plots around $c \sim 0$. This effect is very visible in the $\gamma \gamma$
and $WW^*$ results displayed in the middle plots. On the other hand, in the $ZZ^*$ case the CMS dijet analysis
is less powerful, so there is a weaker suppression of the likelihood around $c \sim 0$.

All the above information is combined in the bottom right panel of Fig.~\ref{fig:ac}, assuming that
there are no virtual non-Standard-Model particles contributing to $H \to \gamma \gamma$ decay
or the $H g g$ coupling. We note that the global fit is not symmetric between the two possibilities for the sign of $c$
relative to $a$, a feature visible in the middle left panel of Fig.~\ref{fig:ac},
and traceable to the interference between the $t$ quark and $W$ boson
loops contributing to the $H \to \gamma \gamma$ decay amplitude. In the past it has been a common
feature of such global fits that they have exhibited two
local minima of the likelihood function with opposite signs of $c$ that, because of this
asymmetry, were not equivalent but had similar likelihoods~\cite{csigndegeneracy}. We see in the bottom right panel of Fig.~\ref{fig:ac},
for the first time a clear preference for the minimum with $c > 0$, i.e., the same sign as
in the Standard Model.

\begin{figure}[h!]
\centering
\includegraphics[scale=0.4]{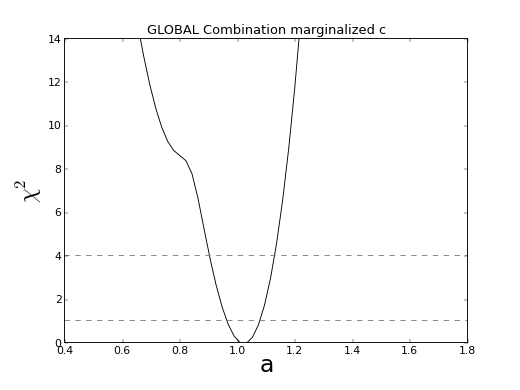}
\includegraphics[scale=0.4]{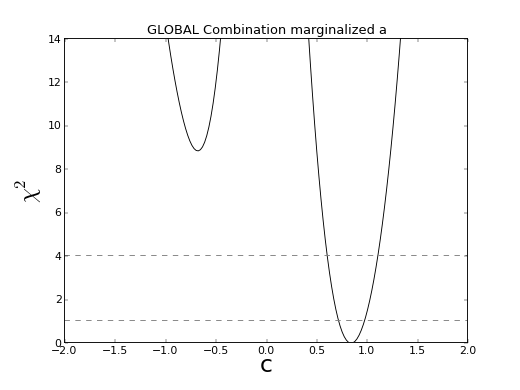}
\caption{\it The one-dimensional likelihood functions for the boson coupling parameter $a$ (left panel) and 
the fermion coupling parameter $c$ (right panel),
as obtained by marginalizing over the other parameter in the bottom right panel of Fig.~\ref{fig:ac}.}
\label{fig:aandc}
\end{figure}

This feature is also seen clearly in Fig.~\ref{fig:aandc}, where we
display in the left panel the one-dimensional likelihood function $\chi^2$ for the boson coupling 
parameter $a$ obtained by marginalizing over the fermion coupling parameter $c$, and in
the right panel the one-dimensional likelihood function for $c$ obtained by marginalizing over $a$.
We see that the fit with $c > 0$ is strongly favoured over that with $c < 0$, with $\Delta \chi^2 \sim 9$. 
The parameters of the global minimum of the $\chi^2$ function
and their 68\% CL ranges are as follows:
\begin{equation}
a \; = \; 1.03 \pm 0.06 \; , \; c \; = \; 0.84 \pm 0.15 \, .
\label{acvalue}
\end{equation}
This preference for $c > 0$ is largely driven by the recently-released CMS $\gamma \gamma$ data.

The yellow lines in the bottom right panel of Fig.~\ref{fig:ac} correspond to various
alternatives to the Standard Model, as discussed above. We see that fermiophobic models
(the horizontal line) are very strongly excluded, as are anti-dilaton models in which $c = -a$. On the
other hand, dilaton/MCHM4 models with $a = c$ are compatible with the data as long as their common value is close
to unity. Likewise, MCHM5 models lying along the curved line are also compatible with the data if
their parameters are chosen to give predictions close to the Standard Model.

\begin{figure}[h!]
\centering
\includegraphics[scale=0.4]{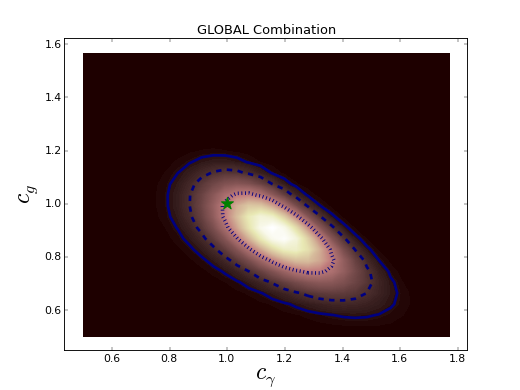}
\includegraphics[scale=0.4]{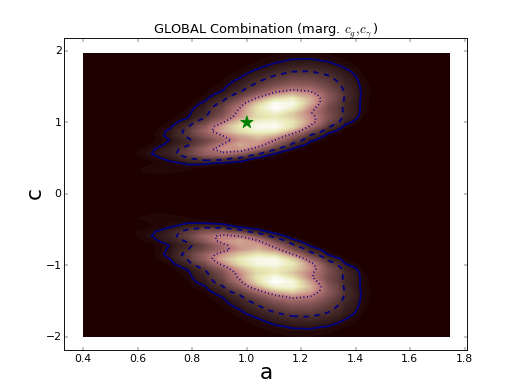}
\caption{\it Left: The constraints in the $(c_\gamma, c_g)$ plane imposed by the measurements in Fig.~\ref{fig:mulist},
assuming the Standard Model values for the tree-level couplings to massive bosons and fermions,
i.e., $a = c = 1$. Right: The constraints in the $(a, c)$ plane when marginalizing over $c_\gamma$ and $c_g$.}
\label{fig:cgammacg}
\end{figure}

The fact that, whereas all the direct measurements of $H$ couplings to fermions and massive vector bosons 
are very compatible with the Standard Model, the coupling to $\gamma \gamma$ was formerly less compatible,
has given rise to much speculation that additional virtual particles may be contributing to the
factor $c_\gamma$ in (\ref{triangles}). However, the motivation for this speculation has been largely
removed by the recent re-evaluation of the $H \to \gamma \gamma$ decay rate by the CMS
Collaboration, which is quite compatible with the Standard
Model prediction. The left panel of Fig.~\ref{fig:cgammacg} shows the results of a global fit to
the anomaly factors $(c_\gamma, c_g)$, assuming the Standard Model values $(a, c) = (1, 1)$
for the tree-level couplings to massive bosons and fermions. Under this hypothesis, 
any deviation from $(c_\gamma, c_g) = (1,1)$ would be due to new particles beyond the Standard Model.
We see explicitly in Fig.~\ref{fig:cgammacg} that, while
there may still be a hint that $c_\gamma > 1$, 
the value of $c_g$ is completely compatible with the
Standard Model. Thus, any set of  new particles contributing to $c_\gamma$ should be
constructed so as not to contribute significantly to $c_g$.

The right panel of Fig.~\ref{fig:cgammacg} is complementary, showing the constraints in the
$(a, c)$ plane after marginalizing over $(c_\gamma, c_g)$. Thus it represents the constraints on
$a$ and $c$ if no assumption is made about the absence of new particle contributions to the
loop amplitudes. In this case, the symmetry between the solutions with $c > 0$ and $< 0$ is restored,
as the $H \to \gamma \gamma$ decay rate no longer discriminates between them. In this case, the
Standard Model values $a = c =1$ are well inside the most favoured region of the $(a, c)$ plane.

\begin{figure}[h!]
\centering
\includegraphics[scale=0.4]{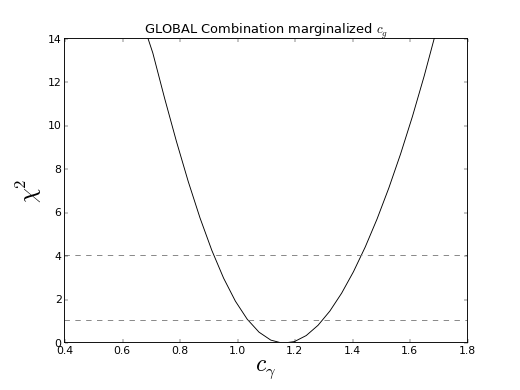}
\includegraphics[scale=0.4]{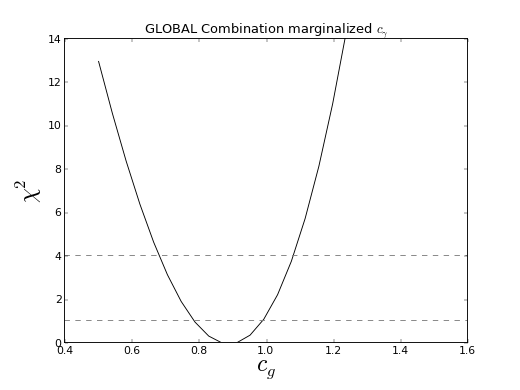}
\caption{\it The one-dimensional likelihood functions for $c_\gamma$ (left panel) and $c_g$ (right panel),
as obtained by marginalizing over the other variable in the bottom right panel of Fig.~\ref{fig:cgammacg},
assuming the Standard Model values for the tree-level couplings to massive bosons and fermions.}
\label{fig:cgammaandcg}
\end{figure}

We display in the left panel of Fig.~\ref{fig:cgammaandcg} the one-dimensional likelihood function 
$\chi^2$ for the factor $c_\gamma$ obtained by marginalizing over $c_g$, and in
the right panel the one-dimensional likelihood function for $c_g$ obtained by marginalizing over $c_\gamma$.
The central values and the 68\% CL ranges of $c_\gamma$ and $c_g$ are as follows:
\begin{equation}
c_\gamma \; = \; 1.18 \pm 0.12 \; , \; c_g \; = \; 0.88 \pm 0.11 \, ,
\label{cgammacgvalues}
\end{equation}
and the likelihood price for $c_\gamma = 1$ is $\Delta \chi^2 = 2$, whereas the price for $c_g = 1$ is $\Delta \chi^2 = 1$.

\section{Probing the Mass Dependence of Higgs Couplings}

\begin{figure}[h!]
\centering
\includegraphics[scale=0.4]{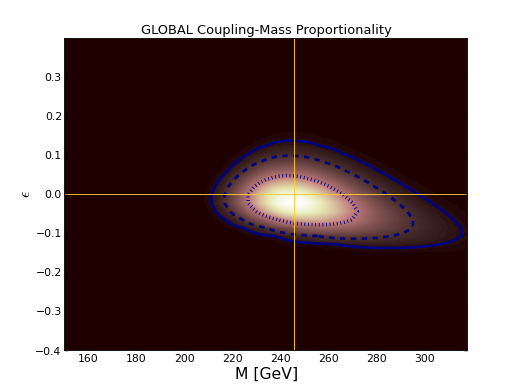}
\includegraphics[scale=0.4]{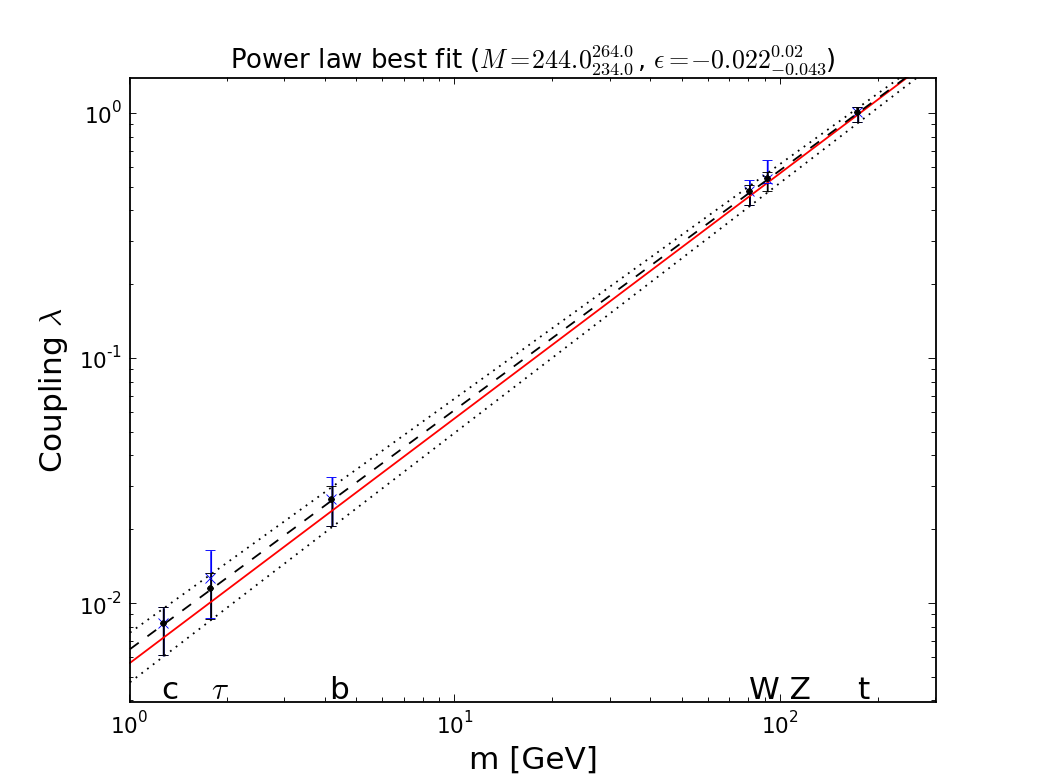}
\caption{\it The constraints in the $(M, \epsilon)$ plane imposed by the measurements in Fig.~\ref{fig:mulist}
(left panel) and the strengths of the couplings to different fermion flavours and massive bosons predicted
by this two-parameter $(M, \epsilon)$ fit (right panel). In the latter, the red line is the Standard Model prediction, the
black dashed line is the best fit, and the dotted lines are the 68\% CL ranges. For each particle species,
the black error bar shows the range predicted by the global fit, and the blue error bar shows the range predicted for
that coupling if its measurement is omitted from the global fit.}
\label{fig:Mepsilon}
\end{figure}

\begin{figure}[h!]
\centering
\includegraphics[scale=0.4]{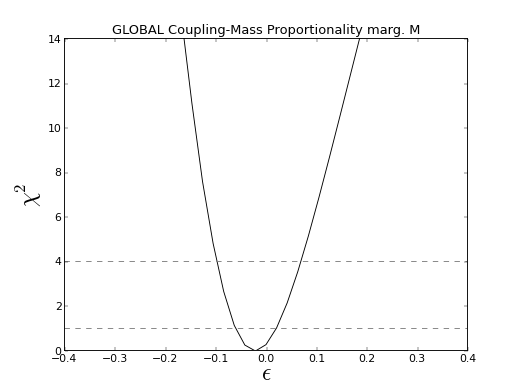}
\includegraphics[scale=0.4]{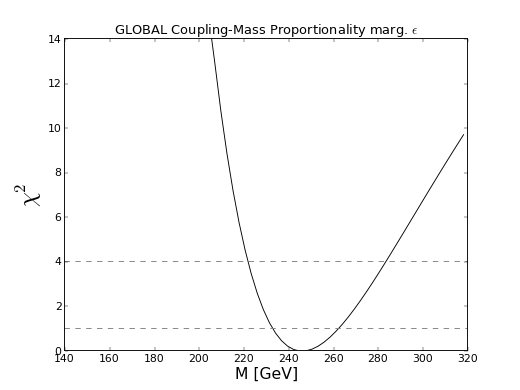}
\caption{\it The one-dimensional likelihood functions for $\epsilon$ (left panel) and $M$ (right panel),
as obtained by marginalizing over the other variable in the left panel of Fig.~\ref{fig:Mepsilon}.}
\label{fig:epsilonandM}
\end{figure}

We now turn to the results of a global fit using the $(M, \epsilon)$ parameterization
(\ref{scalingcouplings}) that probes directly the extent to which the current measurements
constrain the $H$ couplings to other particles to be approximately linear: $\epsilon \sim 0$,
and the extent to which the mass scaling parameter $M \sim v$. The left panel of
Fig.~\ref{fig:Mepsilon} shows the result of combining the measurements shown in
Fig.~\ref{fig:mulist} in the $(M, \epsilon)$ plane. The horizontal and vertical yellow lines correspond to
$\epsilon = 0$ and $M = v$, respectively, and the data are quite compatible with these values.
The central values and the 68\% CL ranges of $M$ and $\epsilon$ are as follows:
\begin{equation}
M \; = \; 244^{+ 20}_{- 10}~{\rm GeV} \; , \; \epsilon \; = \; -0.022^{+ 0.042}_{- 0.021} \, ,
\label{Mepsilonvalues}
\end{equation}
and the likelihood price for $M = 246$~GeV and $\epsilon = 0$ is $\Delta \chi^2 = 0.12$.
It is remarkable that the data already constrain the mass dependence of the $H$ couplings
to other particles to be linear in their masses to within a few \%, and that the mass
scaling parameter $M$ is within 10\% of the Standard Model value $v = 246$~GeV.
We display in the left panel of Fig.~\ref{fig:epsilonandM} the one-dimensional likelihood function 
$\chi^2$ for the factor $\epsilon$ obtained by marginalizing over $M$, and in
the right panel the one-dimensional likelihood function for $M$ obtained by marginalizing over $\epsilon$.

The right panel of Fig.~\ref{fig:Mepsilon} displays the mass dependence of the $H$ couplings in a different way, 
exhibiting explicitly the constraints on the couplings of $H$ to other particles within the 
parameterization (\ref{scalingcouplings}). The solid red line is the prediction of the Standard Model,
$\epsilon = 0$ and $M = v$, the black dashed line corresponds to the best-fit values in
(\ref{Mepsilonvalues}), and the dotted lines correspond to their 68\% CL ranges. The black
points and vertical error bars are the predictions of the $(M, \epsilon)$ fit for the couplings of $H$
to each of the other particle species: the points lie on the best-fit dashed line and the error bars
end on the upper and lower dotted lines. Also shown (in blue) for each particle species is the prediction
for its coupling to $H$ if the data on that particular species are omitted from the global fit. In other words,
the blue points and error bars represent the predictions for the $H$ coupling to that particle, as derived
from the couplings to other particles.

\section{The Total Higgs Decay Rate}

\begin{figure}[h!]
\centering
\includegraphics[scale=0.4]{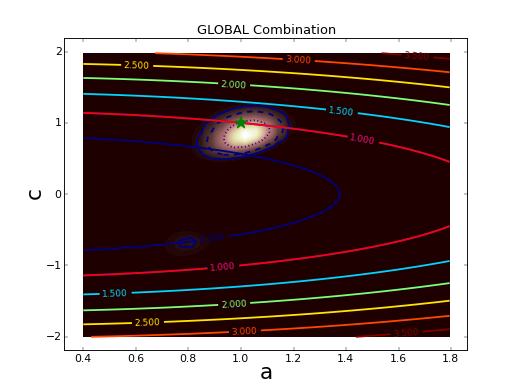}
\includegraphics[scale=0.4]{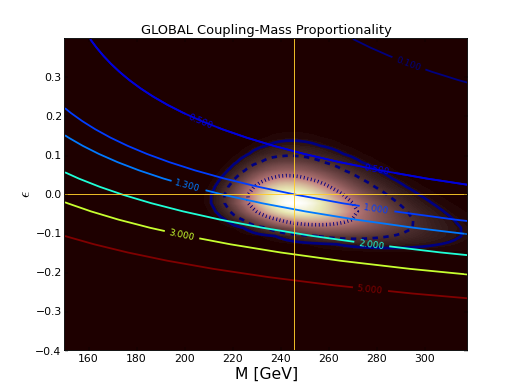}
\caption{\it Contours of the total Higgs decay rate relative to the Standard Model prediction
in the $(a, c)$ plane shown in the bottom right panel of Fig~\ref{fig:ac} (left) and the $(M, \epsilon)$ plane 
shown in the left panel of Fig.~\ref{fig:Mepsilon} (right).}
\label{fig:Rtotal}
\end{figure}

\begin{figure}[h!]
\centering
\includegraphics[scale=0.5]{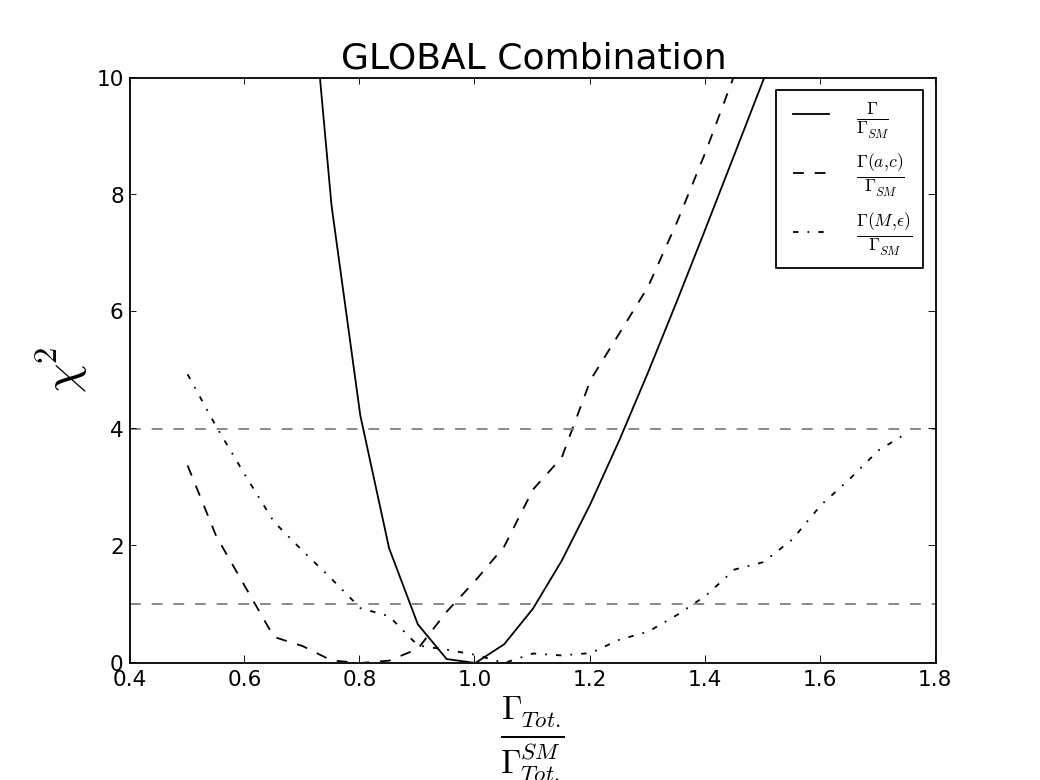}
\caption{\it The one-dimensional likelihood function for the
total Higgs decay width relative to its value in the Standard Model, $R \equiv
\Gamma/\Gamma_{SM}$, assuming decays into Standard Model particles alone and
assuming $a = c$ or equivalently $\epsilon = 0$ (solid line), marginalizing over $(a, c)$
(dashed line) and marginalizing over $(M, \epsilon)$ (dot-dashed line).}
\label{fig:Rtotalmarginalized}
\end{figure}

We now discuss the total Higgs decay rate in the two classes of global fit discussed above,
assuming that the Higgs has no other decays beyond those in the Standard Model \cite{totalwidth}. The
left panel of Fig.~\ref{fig:Rtotal} displays contours of the Higgs decay rate relative to the
Standard Model prediction in the $(a, c)$ plane discussed in Section~3. The local $\chi^2$
minimum with $c > 0$ corresponds to a Higgs decay rate very close to the Standard Model
value, whereas the disfavoured `echo' solution with $c < 0$ has a somewhat smaller decay rate.
The right panel of Fig.~\ref{fig:Rtotal} displays contours of the Higgs decay rate in the
$(M, \epsilon)$ plane, where we again see that the best fit has a total decay rate very
close to the Standard Model value. We display in Fig.~\ref{fig:Rtotalmarginalized}
the one-dimensional likelihood function for the total Higgs decay width
relative to its Standard Model value assuming no contributions from non-Standard-Model
particles. The solid line is obtained assuming that $a = c$ (or, equivalently, that $\epsilon = 0$
but $M$ is free), the dashed line is obtained marginalizing over $(a, c)$, and the dot-dashed line
is obtained by marginalizing over $(M, \epsilon)$.

One may also use the current Higgs measurements to constrain the branching ratio for
Higgs decays into invisible particles, $BR_{inv}$ \cite{invisiblehiggsdecay}. This invisible branching ratio factors out of the total decay width as
\begin{equation}
\Gamma_\text{Tot} = \Gamma_\text{Vis} + \Gamma_\text{Inv} = \left( \frac{R_\text{Vis}}{1- BR_\text{Inv}} \right) \Gamma^\text{SM}_\text{Tot}	\quad ,
\end{equation}
where $R_\text{Vis} = \Gamma_\text{Vis} / \Gamma^\text{SM}_\text{Tot}$ is the rescaling factor of the total decay width in the 
absence of an invisible contribution. Thus we see that an invisible branching ratio acts as a general suppression of 
all other branching ratios, which could be compensated by non-standard visible Higgs decays.

\begin{figure}[h!]
\centering
\includegraphics[scale=0.4]{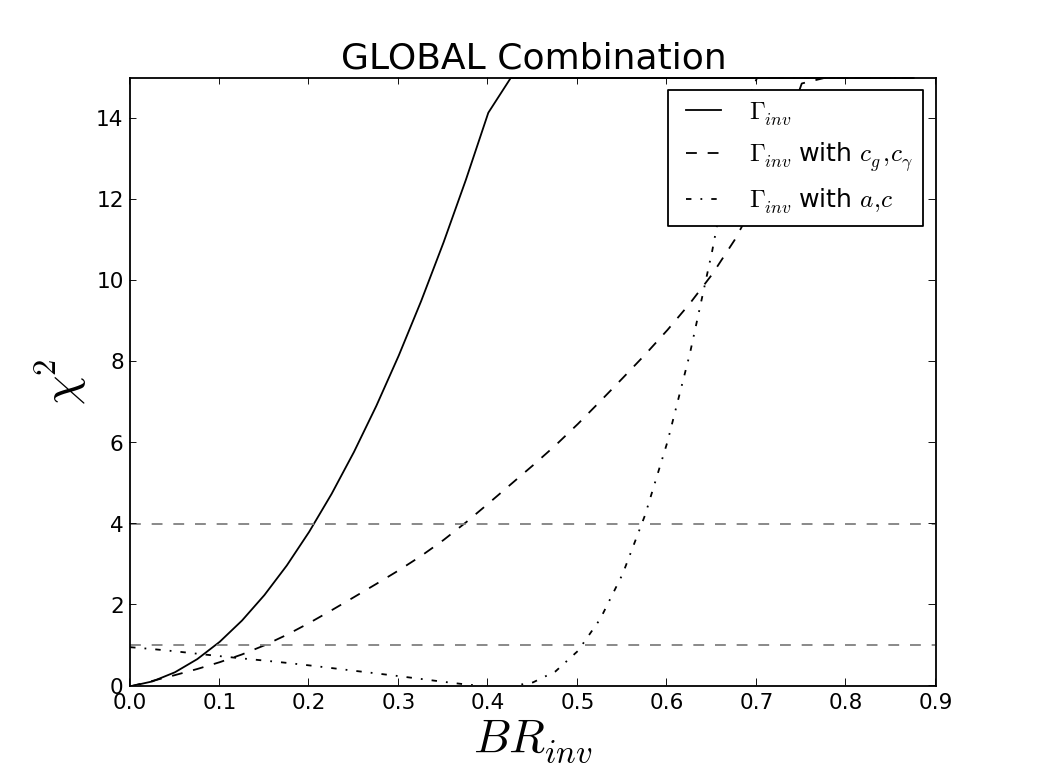}
\includegraphics[scale=0.4]{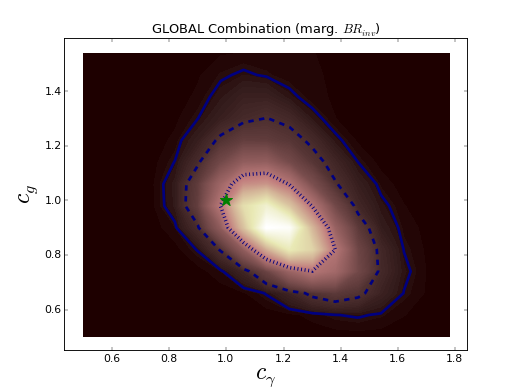}
\caption{\it Left: The branching ratio for Higgs decay into invisible particles obtained assuming the
Standard Model decay rates for all the visible Higgs decays (solid), marginalizing over $(c_\gamma,c_g)$ (dashed) and (a,c) (dot-dashed). Right: The constraints in the 
$(c_\gamma, c_g)$ plane when marginalizing over the invisible branching ration $BR_{inv}$. }
\label{fig:invBR}
\end{figure}

The left panel of Fig.~\ref{fig:invBR} displays the $\chi^2$ function
for $BR_{inv}$ under various assumptions. The solid line was obtained assuming the Standard Model 
couplings for visible particles, i.e., $(a, c) = (1, 1)$
or equivalently $(M, \epsilon) = (v, 0)$. We see that the best fit has $BR_{inv} = 0$, and that the
68 and 95\% CL limits are 0.04 and 0.13, respectively. The dot-dashed line was obtained by
marginalizing over $(a, c)$, where the shallow minimum at $BR_{inv} \sim 0.4$ would require $a>1$. 
Finally, the dashed line was obtained fixing $(a, c) = (1, 1)$ (or equivalently
$(M, \epsilon) = (v, 0)$), but marginalizing over the loop factors $(c_\gamma, c_g)$. Conversely,
the right panel of Fig.~\ref{fig:invBR} displays the constraint in the $(c_\gamma, c_g)$ plane obtained by
marginalizing over $BR_{inv}$.

\section{Conclusions}

The recent installments of data from the LHC experiments announced in March 2013
impose strong new constraints on the properties and couplings of the $H$ particle,
which is beyond doubt a Higgs boson. The data now constrain this particle to have
couplings that differ by only some \% from those of the Higgs boson of the Standard Model.
In particular, the relative sign of its couplings to bosons and fermions
is fixed for the first time, its couplings to other particles are very close to being linear in their masses,
and strong upper limits on invisible Higgs decays can be derived.

The data now impose severe constraints on composite alternatives to the elementary
Higgs boson of the Standard Model. However, they do not yet challenge the predictions of
supersymmetric models, which typically make predictions much closer to the Standard
Model values. We therefore infer that the Higgs coupling measurements, as well
as its mass, provide circumstantial support to supersymmetry as opposed to these
minimal composite alternatives, though this inference is not conclusive.

It is likely that the first LHC run at 7 and 8 TeV has now yielded most of its Higgs secrets,
and we look forward to the next LHC run at higher energy, and its later runs at
significantly higher luminosity. These will provide significant new information about
the $H$ particle and constrain further its couplings, as well as providing opportunities
to probe directly for other new physics. The LHC will be a hard act to follow.

\section*{Acknowledgements}

The work of JE was supported partly by the London
Centre for Terauniverse Studies (LCTS), using funding from the European
Research Council via the Advanced Investigator Grant 267352.
The work of TY was supported by a Graduate Teaching Assistantship from
King's College London. JE thanks CERN for kind hospitality.

\end{document}